\documentclass{aip-cp}

\usepackage[numbers]{natbib}
\usepackage{rotating}
\usepackage{graphicx}

\begin{document}

\title{Covariant energy density functionals: the assessment of
global performance across the nuclear landscape}

\author[aff1]{A.\ V.\ Afanasjev\corref{cor1}}

\affil[aff1]{Department of Physics and Astronomy, Mississippi State
University, Mississippi State, MS 39762, USA}
\corresp[cor1]{Corresponding author: Anatoli.Afanasjev@gmail.com}

\maketitle

\begin{abstract}
  The assessment of the global performance of the state-of-the-art covariant 
energy density functionals  and related theoretical uncertainties in the 
description of ground state observables has recently been performed. 
Based on these results, the correlations between global description of
binding energies and nuclear matter properties of covariant energy density 
functionals have been studied in this contribution.
\end{abstract}

\section{INTRODUCTION}

   The global performance of the covariant energy density functionals (CEDF's)
in the description of ground state observables has been assessed in Refs.\ 
\cite{AARR.13,AARR.14,AARR.15} employing the state-of-the-art functionals 
NL3*, DD-ME2, DD-ME$\delta$, and DD-PC1. They represent three classes of 
functionals  which differ by basic model assumptions and fitting protocols.  
The available experimental data on ground state properties of even-even 
nuclei such as binding  energies, charge radii, neutron skin thicknesses 
as well as two-proton and two-neutron separation energies and the positions
of two-proton and two-neutron drip lines have been confronted with the 
results of the calculations. For the first time, theoretical systematic 
uncertainties in the prediction of physical observables have been investigated 
on a global scale for relativistic functionals. Special attention has been 
paid to the propagation of these uncertainties towards the neutron-drip line 
\cite{AARR.14} and the sources of these uncertainties \cite{AARR.15}. Since
the details of these studies are easily accessible, I will focus in this
contribution on the relations between global description of masses and
nuclear matter properties of underlying functionals.

\section{THE COMPARISON OF DIFFERENT FUNCTIONALS}

\begin{figure}[h]
\begin{tabular}{c}
\includegraphics[width=370pt]{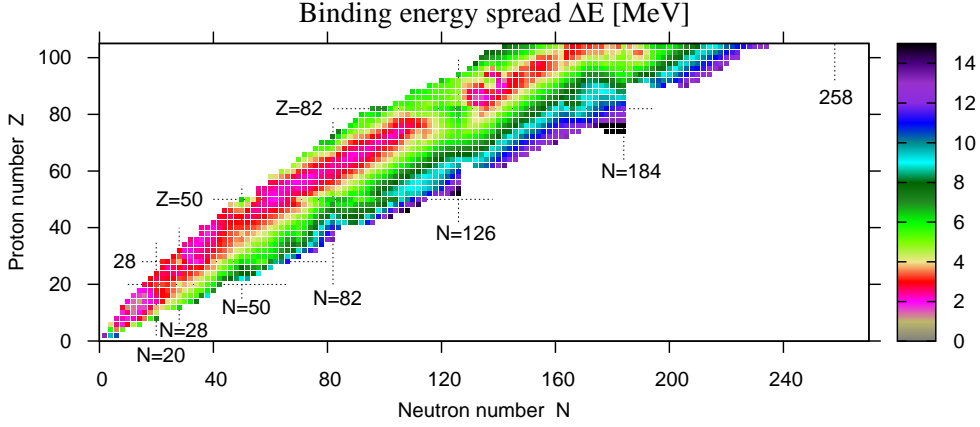} 
\end{tabular}
\caption{The binding energy spreads $\Delta E(Z,N)$ as a function 
of proton and neutron number. $\Delta E(Z,N) = |E_{max}(Z,N) - E_{min}(Z,N)|$, 
where $E_{max}(Z,N)$ and $E_{min}(Z,N)$ are the largest and the smallest 
binding energies for each $(N,Z)$ nucleus obtained with the NL3*, DD-ME2,
DD-ME$\delta$ and DD-PC1 functionals. From Ref.\ \cite{AARR.14}.}
\label{Fig.1}
\end{figure}

  In Fig.\ \ref{Fig.1} the map of theoretical uncertainties $\Delta E(Z,N)$ in 
the description of binding energies is shown. The comparison of this figure with 
Fig.\ 1 in Ref.\ \cite{AARR.13} (which presents experimentally known nuclei in 
the nuclear chart), shows that the spreads in the predictions of binding energies 
stay within 5-6 MeV for the known nuclei. These spreads are even smaller (typically 
around 3 MeV) for the nuclei in the valley of beta-stability. However, theoretical
systematic uncertainties for the masses increase drastically when approaching
the neutron-drip line and in some nuclei they reach 15 MeV. This is a
consequence of poorly defined isovector properties of many CEDF's.

\begin{figure}[ht]
\begin{tabular}{c}
\includegraphics[width=345pt]{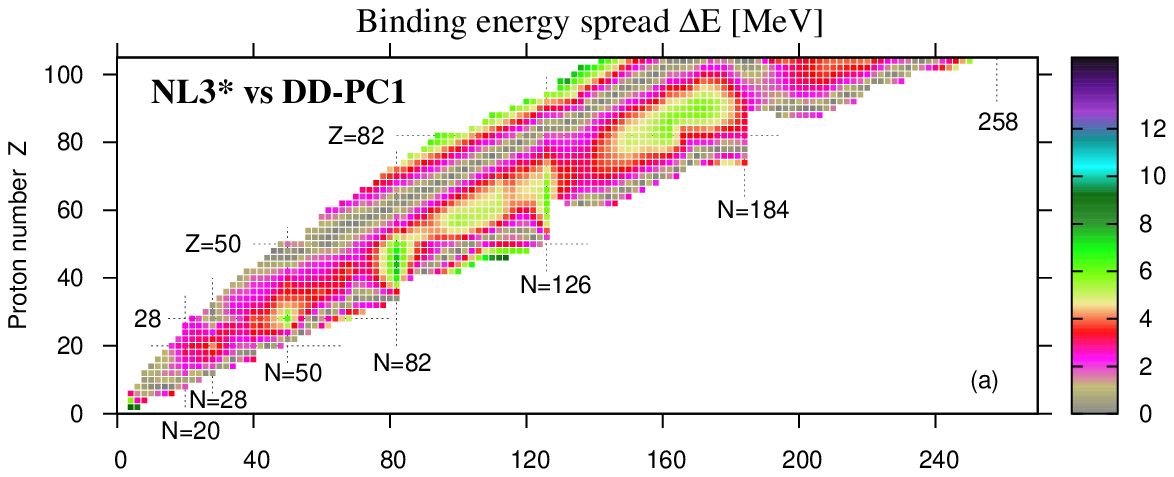} \\
\includegraphics[width=345pt]{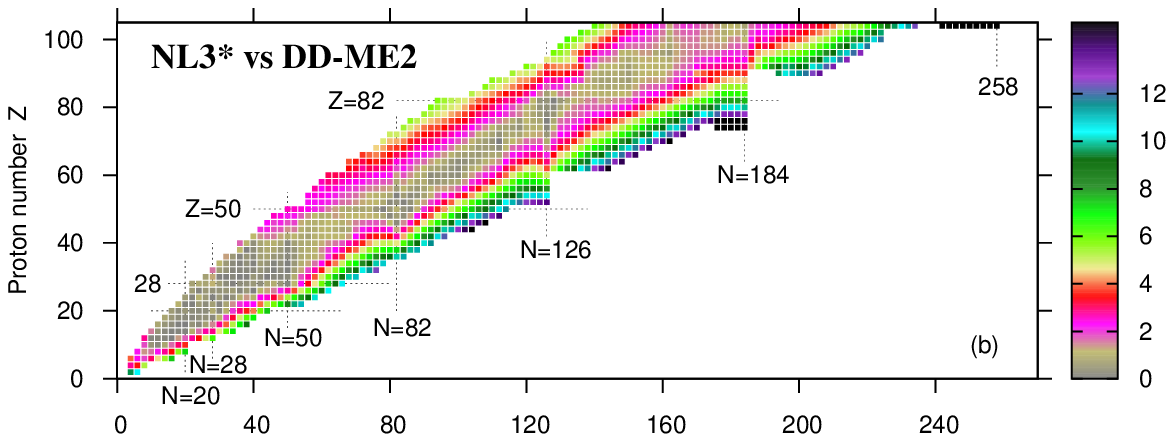} \\ 
\includegraphics[width=345pt]{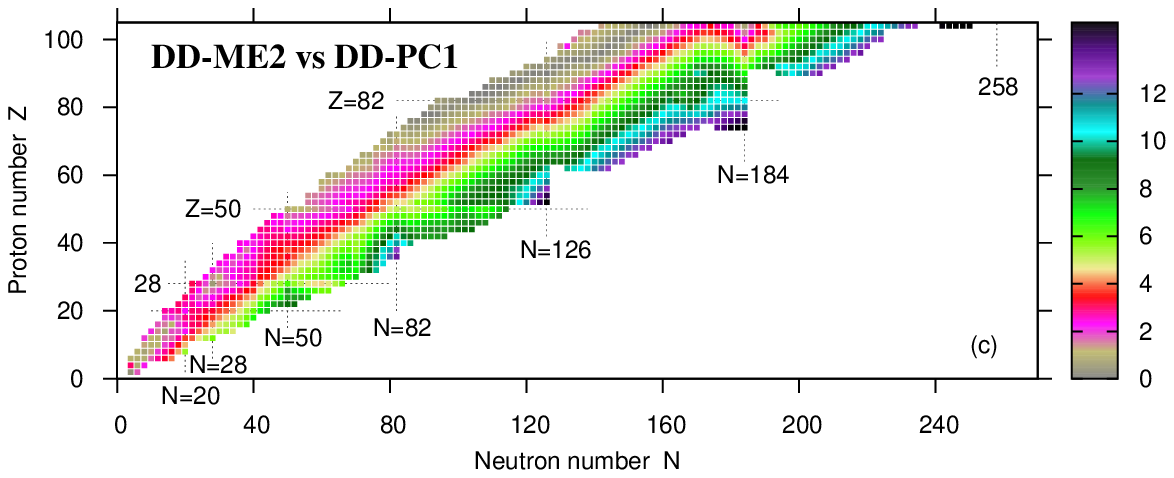} \\
\end{tabular}
\caption{The same as Fig.\ \ref{Fig.1}, but for binding energy spreads 
$\Delta E(Z,N)$ of the pairs of indicated functionals.}
\label{Fig.2}
\end{figure}

   The fitting protocols of employed functionals always contain data on 
finite nuclei and nuclear matter properties (see Sect.\ II in 
Ref.\ \cite{AARR.14} for more details). The data on finite nuclei 
includes binding energies, charge radii and occasionally neutron skin 
thicknesses. The isovector properties of the functionals are affected 
by their nuclear matter properties. However, it is not always possible
to find one-to-one correspondence between the differences in nuclear 
matter properties of two functionals and the differences in their 
description of masses.  This can be illustrated by the comparison of binding 
energy spreads for the pairs of the functionals (Fig.\ \ref{Fig.2}) with 
their nuclear matter properties (Table 1). The smallest difference in the 
predictions of binding energies exists for the DD-ME2/DD-ME$\delta$ pair of 
the functionals (see Fig.\ 9 in Ref.\ \cite{AARR.14}); for almost half of 
the $Z\leq 104$ nuclear landscape their predictions differ by less than 
1.5 MeV and only in a few points of nuclear landscape the difference in 
binding energies of two functionals exceeds 5 MeV. The nuclear matter 
properties of these two functionals are similar with some minor differences 
existing only for the incompressibility $K_{\infty}$ and Lorentz effective mass 
$m$*/$m$ (Table 1). However, there is an opposite example of the pair
of the DD-ME2 and DD-PC1 functionals for which substantial differences in the 
mass predictions (Fig.\ \ref{Fig.2}c) exist for quite similar nuclear matter 
properties (Table 1). Note that these differences in mass predictions
are larger than the ones 
for the NL3*/DD-PC1 pair of the functionals (Fig.\ \ref{Fig.2}a) which are 
characterized by a substantial difference in the energy per particle ($E/A$), 
symmetry energy $J$ and its slope $L$ (Table 1).

  It is clear that the part of the difference in mass predictions is coming 
from the use of different data on finite nuclei in fitting protocols; the 
binding energies of these nuclei provide the normalization of the energy for 
the functional. The most similar fitting protocols exist
in the case of the NL3* and DD-ME2 functionals which were fitted to the
same 12 spherical nuclei \cite{NL3*,DD-ME2}. The comparison of Figs. 1a,b  of Ref.\ 
\cite{AARR.14} with Fig.\ \ref{Fig.2}b in the present manuscript 
clearly illustrates that the difference in binding energies is minimal 
for these two functionals in the gray band region of Fig.\ \ref{Fig.2}b which 
overlaps with the nuclei used in the fitting protocol. 
Fig.\ \ref{Fig.2} also illustrate the fact that most rapid increase of
the differences in the predicted binding energies takes place not in 
the direction of isospin but in the direction which is perpendicular 
to the gray band of similar energies. These differences are due in part
to different nuclear matter properties of these two functionals.

\begin{table*}[ht]
\caption{ Properties of symmetric nuclear matter at saturation for the covariant energy density 
functionals: the density $\rho_0$, the energy per particle $E/A$, the incompressibility 
$K_{\infty}$, the symmetry energy $J$ and its slope $L$, and the Lorentz effective mass $m$*/$m$
\cite{JM.89} of a nucleon at the Fermi  surface. The quantities which are located beyond the
limits of the SET2b constraint set of Ref.\ \cite{RMF-nm} are shown in bold. The last column
shows the rms deviations $\Delta E_{rms}$ between calculated and experimental binding energies.
For first four functionals, they are defined in Ref.\ \cite{AARR.14} with respect of 640 
measured masses presented in the AME2012 compilation \cite{AME2012}. For PC-PK1 they are defined 
with respect of 575 masses in Ref.\ \cite{ZNLYM.14}.
}
\label{tab-nuclear-matter}
\begin{center}
\begin{tabular}{|c|c|c|c|c|c|c|c|c|}\hline
CEDF              & $\rho_0$ [fm$^{-3}$] & $E/A$ [MeV] & $K_{\infty}$ [MeV] & $J$ [MeV] & $L$ [MeV] &  $m$*/$m$ & & $\Delta E_{rms}$ [MeV] \\ \hline
      1                &       2        &       3           &   4    &            5  &    6         &   7    &  & 8     \\ \hline
NL3* \cite{NL3*}       &   0.150        &  -16.31           & 258    & {\bf 38.68}   & {\bf 122.6}  &  0.67  &  & 2.96  \\
DD-ME2 \cite{DD-ME2}   &  0.152         & -16.14            & 251    & 32.40         &  49.4        & 0.66   &  & 2.39  \\
DD-ME$\delta$ \cite{DD-MEdelta} & 0.152 & -16.12            & 219    & 32.35         &  52.9        & 0.61   &  & 2.29  \\
DD-PC1 \cite{DD-PC1,PC-PK1}&   0.152    & -16.06            & 230    & 33.00         &  68.4        & 0.66   &  & 2.01  \\
PC-PK1 \cite{PC-PK1}   &   0.154        & -16.12            & 238    & {\bf 35.6}    &  {\bf 113}   & 0.65   &  & 2.58   \\ \hline
\end{tabular}
\end{center}
\end{table*}

\section{NUCLEAR MATTER CONSTRAINTS}

  The question to which extent nuclear matter constraints are important
(and how strictly they have to be imposed) for the definition of the 
properties of covariant energy density functionals still remains not
fully answered. Definitely, the equation of state (EOS) relating
pressure, energy density, and temperature at a given particle number density
is essential for modeling neutron stars, core-collapse supernovae, mergers 
of neutrons stars and the processes (such as nucleosynthesis) taking places 
in these environments. However, the properties of finite nuclei are in addition 
defined by the underlying shell structure which depends sensitively on the 
single-particle features \cite{VALR.05}. 

   Recent analysis of the 263 relativistic functionals with respect of nuclear 
matter constraints has been performed in Ref.\ \cite{RMF-nm}. Note that only around 
ten of these functionals have been used in a more or less systematic studies of the 
properties of finite nuclei; the performance of other functionals with respect of 
the description of finite nuclei (apart of few spherical nuclei used in the fitting
protocols) is not known. Three different sets of constraints related to 
symmetric nuclear matter, pure neutron matter, symmetry energy and its derivatives were 
employed in the analysis of Ref.\ \cite{RMF-nm}. Among these 263 functionals only 4 
and 3 satisfy nuclear matter constraint sets called SET2a and SET2b, respectively.
However, these functionals have never been used in the studies of finite nuclei.
Thus, it is impossible to verify whether good nuclear matter properties of the
functional will translate into good global description of binding energies, charge radii, 
deformations etc. Removing isospin incompressibility constraint increases the number of 
functionals which satisfy SET2a and SET2b constraints to 35 and 30, respectively 
\cite{RMF-nm}. Again the performance of absolute majority of these functionals in 
finite nuclei is not known. However, among those are the FSUGold and DD-ME$\delta$ 
covariant energy density 
functionals the global performance of which has been studied in the RMF+BCS and RHB 
models in Refs.\ \cite{RA.11,AARR.14}, respectively. FSUGold is characterized by the 
largest rms deviations from experiment for binding energies (6.5 MeV) among all CEDF's 
the global performance of which is known \cite{AARR.14}. Although the DD-ME$\delta$ 
functional provides quite reasonable description of the binding energies (Table 1), 
it generates unrealistically low inner fission barriers in superheavy elements
\cite{AANR.15}.

  The analysis of Refs.\ \cite{AARR.14,AAR.10,AO.13,AANR.15} clearly indicates that
the NL3*, DD-ME2, PC-PK1 and DD-PC1 CEDF's represent better and well-rounded 
functionals as compared with FSUGold and DD-ME$\delta$. They are able to describe not 
only ground state properties but also the properties of excited states. This is despite 
the fact that first three functionals definitely  fail to describe some of the nuclear 
matter properties (see Table 1 and Ref.\ \cite{RMF-nm}). It is not clear whether that is also
a case for DD-PC1 since it was not analyzed in Ref.\ \cite{RMF-nm}. As a result, one can 
conclude  that the functionals, which provide good nuclear matter properties, do not 
necessary well describe finite nuclei. Such a possibility has already been mentioned in 
Ref.\ \cite{RMF-nm}.

\section{CONCLUSIONS}

The correlations between global description of masses and nuclear matter properties
of the underlying functionals has been discussed based on the results of recent 
assessment of global performance of covariant energy density functionals presented
in Refs.\ \cite{AARR.13,AARR.14,AARR.15}.
It was concluded that the strict enforcement of the limits on the nuclear 
matter properties defined in Ref.\ \cite{RMF-nm} will not necessary (i) lead to the 
functionals with good description of masses and (ii) substantially decrease the 
uncertainties in the description of masses in neutron-rich systems. This is quite 
likely related to the mismatch of phenomenological content,  existing in all modern 
functionals, related to nuclear matter physics and the physics of finite nuclei; the 
later being strongly affected by underlying shell effects.

\section{ACKNOWLEDGMENTS}

  This material is based upon work supported by the U.S. Department of Energy, 
Office of Science, Office of Nuclear Physics under Award Number DE-SC0013037. 
I would like to express my deep graditude to my collaborators, S.\ E.\ Agbemava,
D. Ray and P. Ring, who contributed to Refs.\ \cite{AARR.13,AARR.14,AARR.15}; 
the numerical results of Ref.\ \cite{AARR.14} were used in the preparation of
Fig. 2 in this contrbution.

\bibliographystyle{aipnum-cp}%
\bibliography{references10-short}%

\end{document}